\documentclass[ 
    ,final            % use final for the camera ready runs 
  ] 
  {aipproc} 
\def\fc{f_{\rm c}}

\layoutstyle{6x9} 
 
\begin{document} 
 
\title{Model atmospheres \\of X-ray bursting neutron stars 
} 
 
 
\author{V. Suleimanov}{ 
  address={Insitute for Astronomy and Astrophysics, Kepler Center for Astro and Particle Physics, Eberhard Karls University, Sand 1, 72076 T\"ubingen, Germany}, 
altaddress={Kazan Federal University, Kremlevskaya str. 18, 42008 Kazan, Russia}, 
email={suleimanov@astro.uni-tuebingen.de} 
} 
 
\author{J. Poutanen}{ 
  address={Astronomy Division, PO Box 3000, FIN-90014 University of Oulu, Finland} 
} 
\author{M. Revnivtsev}{ 
  address={Space Research Institute, Profsoyuznaya str. 84/32, 117997 Moscow, Russia} 
} 
\author{K. Werner}{ 
  address={Insitute for Astronomy and Astrophysics, Kepler Center for Astro and Particle Physics, Eberhard Karls University, Sand 1, 72076 T\"ubingen, Germany}, 
} 
 
\begin{abstract} 
We present an extended set of model atmospheres and emergent spectra 
of X-ray bursting neutron stars in low mass X-ray binaries. Compton 
scattering is taken into account. The models were computed in LTE 
approximation for six different chemical compositions: pure hydrogen and pure 
helium atmospheres, and atmospheres 
with a solar mix of hydrogen and helium and various heavy elements 
abundances; $Z$ = 1, 0.3, 0.1, and 0.01 $Z_\odot$, for three values of gravity, $\log g$ =14.0, 14.3, and 14.6 and for 
20 values of relative luminosity 
$l = L/L_{\rm Edd}$ in the range 0.001 - 0.98. The emergent spectra of all models are 
fitted by  diluted blackbody spectra in the observed {\it RXTE}/PCA band 3 - 20 keV and the 
corresponding values of color correction factors $\fc$ are 
presented. We also show 
how to 
use these dependencies to estimate the neutron star's basic parameters.   
 
\end{abstract} 
 
\maketitle 
 
%%%%%%%%%%%%%%%%%%%%%%%%%%%%%%%%%%%%%%%%%%%% 
%% MAINMATTER 
%%%%%%%%%%%%%%%%%%%%%%%%%%%%%%%%%%%%%%%%%%%% 
 
%\section{Method, results of modeling, and application} 
 
The most important and useful sources for the aim of neutron stars (NSs)  $M$ and $R$ finding 
are X-ray bursting NSs with photospheric radius expansion \cite{LvPT:93}. 
 The relation between observed normalization $K$ (for blackbody fit of spectra) and the 
real ratio of NS radius $R$ to distance on late outburst phases is: 
\begin{equation} \label{u1} 
   K^{1/2} = \frac{R_{\rm BB}{\rm (km)}}{D_{10}} = \frac{R{\rm (km)}}{\fc^2~D_{10}}(1+z), 
\end{equation} 
where $D_{10}$ is the distance in units of10 kpc, and $\fc = T_{\rm c}/T_{\rm eff}$ is a color correction factor. 
Therefore, on these phases $K(t)$ dependence reflects $\fc(t)$ dependence only.  
We suggest to fit the observed $K^{-1/4}$ -- $F$ relation by a theoretical $\fc$ -- $l\equiv L/L_{\rm Edd}$ relation, where $F$ is  
the integral observed flux. From this fit we can obtain two independent values:  $R{\rm (km)}\times(1+z)/D_{10}$ 
and $F_{\rm Edd} \sim L_{\rm Edd}/((1+z)D^2_{10})$. Combining these values, we can obtain an observed $M/R$ relation, 
which is independent on the distance and physically corresponds to a maximum possible effective temperature on  
the NS surface. 
If the distance is known  
 we can find $M$ and $R$ 
simultaneously. 
For this method extended theoretical $\fc(l)$ calculations are necessary. 
 
We computed model 
atmospheres of X-ray bursting NSs subject to the constraints 
of hydrostatic and radiative equilibrium assuming planar geometry 
in LTE approximation with Compton scattering taken into account (see details  
of the code in \cite{SP:06,SW:07}). 
 
\begin{figure} 
  \includegraphics[height=.185\textheight]{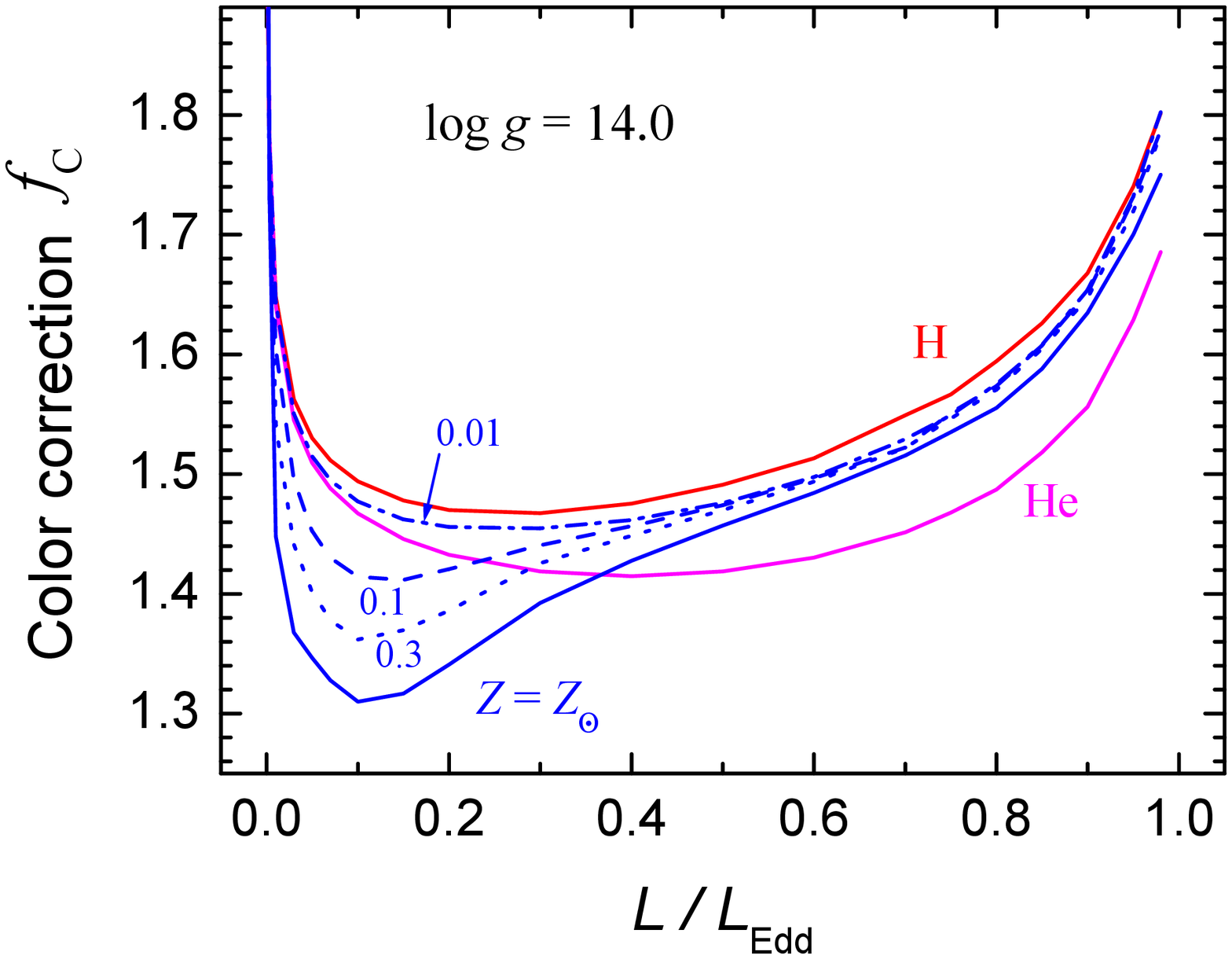} 
  \includegraphics[height=.175\textheight]{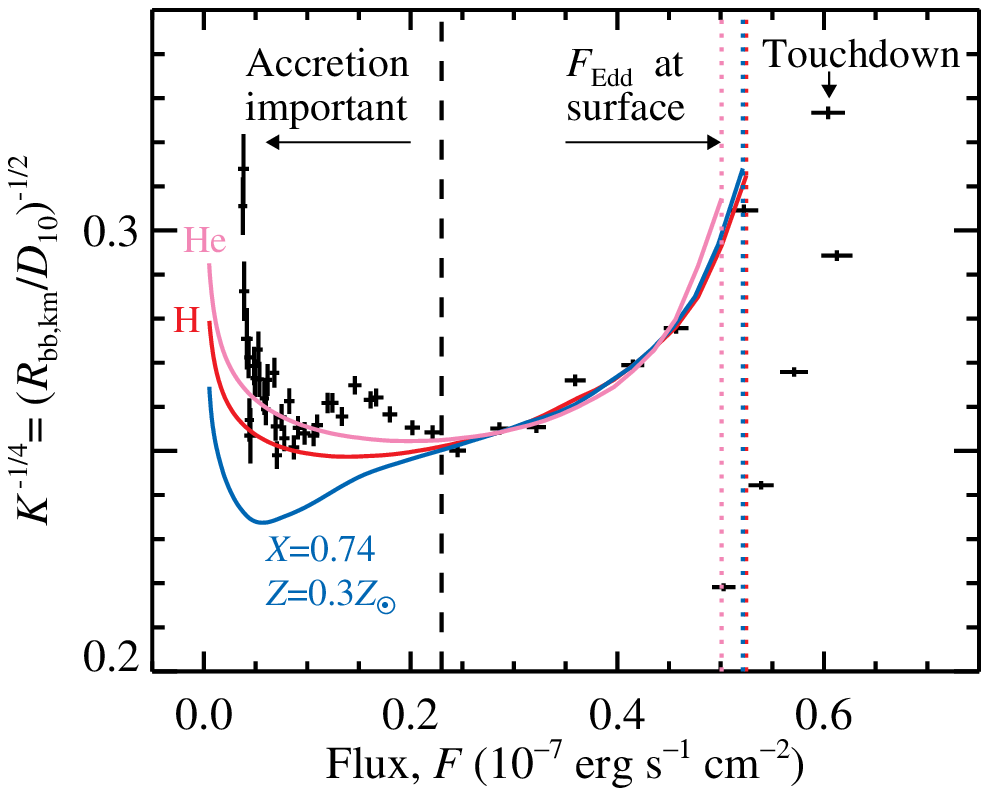} 
  \includegraphics[height=.18\textheight]{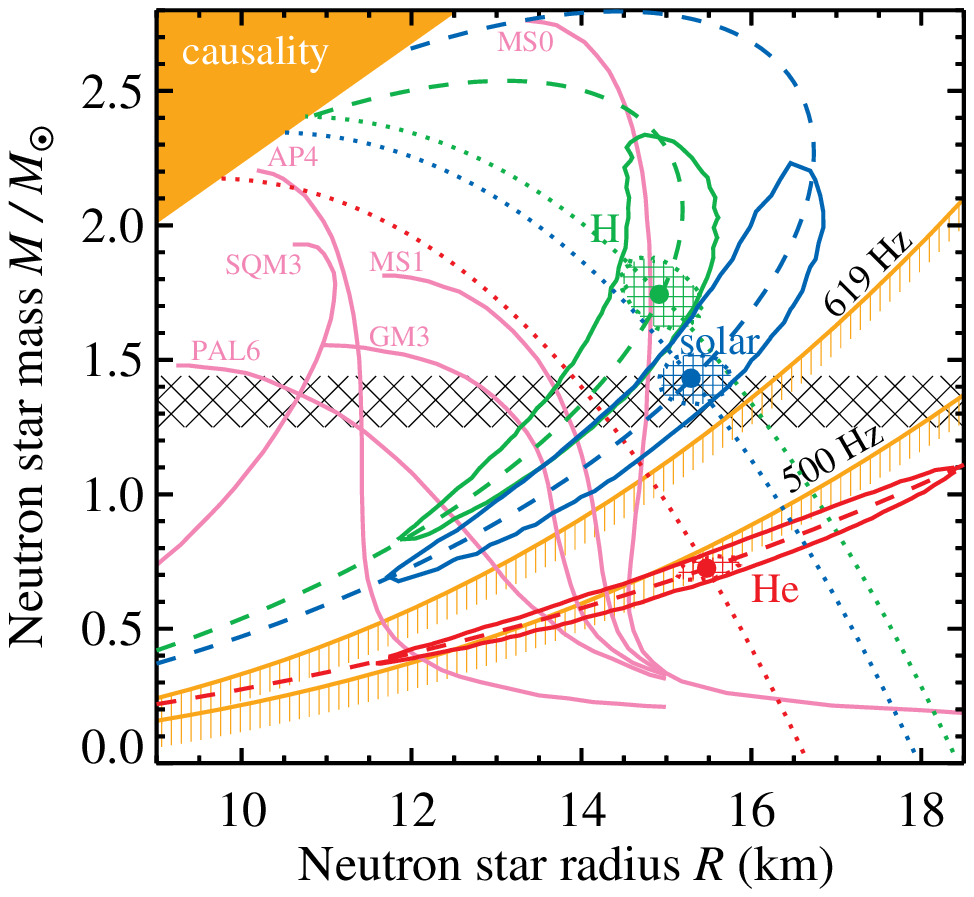} 
  \caption{ \label{v1sfig1} 
{\it Left:} Dependence of color correction factors on the relative luminosity  
for low gravity and various chemical compositions in NS atmosphere models. 
 {\it Middle:} Comparison of the observed dependence of $K^{-1/4} - F$ for 4U\,1724$-$307 (croses) to the best fit theoretical models 
$\fc - l$.  
 {\it Right:} Constraints on  mass and radius of the neutron star 4U\,1724$-$307.  
} 
\end{figure}

We calculated an extended set of NS model 
atmospheres with 6 chemical compositions (pure H, He, and solar H/He mixture 
with $Z$ = 1, 0.3, 0.1 and 0.01 $Z_\odot$), 3 surface gravities: $\log~g$ = 14.0, 14.3 and 
14.6, and 20 luminosities $L$: 0.001, 0.003, 0.01, 0.03, 0.05, 0.07, 0.1, 0.15, 0.2, 0.3, 0.4, 
0.5, 0.6, 0.7, 0.75, 0.8, 0.85, 0.9, 0.95, and 0.98 $L_{\rm Edd}$. Corresponding $T_{\rm eff}$ were 
calculated from $L$ using $\log g$ and chemical composition. 
The model emergent redshifted spectra were fitted 
by diluted blackbody spectra $F_{\rm E} = wB_{\rm E}(\fc T_{\rm eff})$  
%using four slightly different procedures 
in the {\it RXTE}/PCA energy band $3 - 20$ keV. Here $w \approx \fc^{-4}$ is the dilution factor. 
The accepted redshifts were calculated from $\log g$ assuming $M = 1.4 M_\odot$. 
Results are partially presented in Fig.\,\ref{v1sfig1}, left panel. 
 
%\section{Application to 4U\,1724$-$307 and conclusions} 
 
 We fitted the observed $K^{-1/4}$ -- $F$ relation, obtained for the extremely long outburst of 
4U\,1724$-$307~  in November 8, 1996 ({\it RXTE}, \cite{molkov:00}) by computed $\fc$ -- $l$ relations. 
We obtained limitations on $R$ and $M$ for the adopted 
distance $D = 5.3 \pm 0.6$ kpc \cite{ort:97} and various chemical compositions, see Fig.\,\ref{v1sfig1}. 
The values of $M$ and $R$ obtained for H-rich atmospheres  correspond to a stiff Equation of State in the inner NS core. 
Helium atmospheres are not acceptable. More details can be found in \cite{SPRW:10}. 
 
\begin{theacknowledgments} 
  
The work is supported by the DFG grant SFB / Transregio 7 ``Gravitational Wave Astronomy'' (V.S.),  
Russian Foundation for Basic Research (grant  09-02-97013-p-povolzhe-a, V.S.), and 
the Academy of Finland (grant 127512, J.P.). 
\end{theacknowledgments}

\bibliography{Suleimanov_Valery1}
\bibliographystyle{aipproc}

 
 
 
%\bibliography{sample} 
 
 
\end{document}